# Switching the Ferroelectric Polarization by External Magnetic Fields in the Spin = 1/2 Chain Cuprate LiCuVO$_4$

F. Schrettle$^1$, S. Krohns$^1$, P. Lunkenheimer$^{1,*}$, J. Hemberger$^{1,\dagger}$, N. Büttgen$^1$, H.-A. Krug von Nidda$^1$, A. V. Prokofiev$^2$, and A. Loidl$^1$

$^1$Experimental Physics V, Center for Electronic Correlations and Magnetism, University of Augsburg, 86135 Augsburg, Germany
$^2$Institut für Festkörperphysik, Technische Universität Wien, 1040 Wien, Austria

We present a detailed study of complex dielectric constant and ferroelectric polarization in multiferroic LiCuVO$_4$ as function of temperature and external magnetic field. In zero external magnetic field, spiral spin order with an **ab** helix and a propagation vector along the crystallographic **b** direction is established, which induces ferroelectric order with spontaneous polarization parallel to **a**. The direction of the helix can be reoriented by an external magnetic field and allows switching of the spontaneous polarization. We find a strong dependence of the absolute value of the polarization for different orientations of the spiral plane. Above 7.5 T, LiCuVO$_4$ reveals collinear spin order and remains paraelectric for all field directions. Thus this system is ideally suited to check the symmetry relations for spiral magnets as predicted theoretically. The strong coupling of ferroelectric and magnetic order is documented and the complex ($B,T$) phase diagram is fully explored.

## I. INTRODUCTION

Recently the detection of multiferroicity in the $S = 1/2$ spin-chain compounds LiCu$_2$O$_2$ (Ref. 1) and LiCuVO$_4$ (Ref. 2,3), linked together two active research areas of condensed matter physics, namely multiferroicity and quantum spin systems. The discovery of ferroelectricity (FE) in spiral magnets has strongly revived the field of multiferroicity (see Refs. 4,5,6,7,8 and references therein). In these compounds complex spin order, which arises from frustrated and competing interactions, is established at low temperatures and induces ferroelectricity. While there is no generally accepted microscopic mechanism for the generation of FE in multiferroics, a number of models providing fundamental and plausible explanations have been developed.[9,10,11,12,13] In most of these systems, e.g., in the rare-earth manganites like TbMnO$_3$, DyMnO$_3$,[14] TbMn$_2$O$_7$,[15] and Eu:YMnO$_3$,[16] or in Ni$_3$V$_2$O$_8$,[17] FE appears in magnetic phases with spiral or helical order. It has been argued[6] that qualitatively these spin structures already break inversion symmetry and FE is induced via spin-orbit coupling. The antisymmetric Dzyaloshinskii-Moriya interaction has been identified as the most likely mechanism prevailing in these materials.[10,11] A microscopic model based on spin currents in non-collinear magnets has been proposed by Katsura et al.[9] Symmetry considerations further reveal that finite polarization $P$ only appears if the vector product of the spiral axis **e** and the propagation vector of spin order **Q** is finite, i.e. $\mathbf{P} \propto \mathbf{e} \times \mathbf{Q}$.[9,10] This specific prediction follows from general symmetry considerations taking into account that magnetism always breaks time inversion, but ferroelectricity only can evolve when spatial symmetry is broken. In this way the suppression of polarization by external magnetic fields and the complex ($B,T$) phase diagram in Ni$_3$V$_2$O$_8$ have been explained in a Landau-like theory for continuous phase transitions.[17]

Frustrated spin-1/2 systems display a rich variety of exotic ground states and have attracted considerable attention during the last decade.[18] The simplest frustrated model systems probably are $S = 1/2$ spin chains with competing nearest ($J_1$) and next-nearest ($J_2$) neighbor interactions. A $T = 0$ K phase diagram for a $S = 1/2$ quantum spin chain with competing $J_1$ and $J_2$ exchange has been calculated by Bursill et al.,[19] resulting in spiral spin order with a pitch angle depending on the ratio of $J_2$ and $J_1$ for a wide range of parameters. Two prominent examples of quantum spin chains, namely LiCu$_2$O$_2$ (Ref. 20) and LiCuVO$_4$,[21,22] indeed reveal spiral spin order at low temperatures. With the onset of the complex magnetic order also electrical polarization emerges in both systems,[1,2,3] which therefore can be assigned as multiferroic $S = 1/2$ quantum spin chains.

In the present report, we study the evolution of ferroelectric polarization in LiCuVO$_4$ as function of external magnetic fields. In this quantum spin chain, the spiral axis can easily be switched[23] by magnetic field and finally the spiral structure can be completely suppressed[23,24], which should allow for significant tests of the above-mentioned symmetry considerations.[9,10] However, recently Moskvin and Drechsler[13] proposed that FE in LiCuVO$_4$ is induced by disorder, which in lowest order generates polarization with **P** || **a** only. In addition, from detailed electron-spin resonance (ESR) experiments[25] it has been concluded that at least in the paramagnetic phase the antisymmetric Dzyaloshinskii-Moriya interaction plays no role in the magnetic exchange of LiCuVO$_4$. Therefore a detailed study of the dielectric properties as function of magnetic field seems necessary to construct realistic microscopic models for spiral magnets.

LiCuVO$_4$ crystallizes within an orthorhombically distorted inverse spinel structure. The non-magnetic V$^{5+}$ ions occupy the tetrahedrally coordinated *A* sites, while Li$^+$ and Cu$^{2+}$ (3$d^9$ configuration, $S$ = 1/2) occupy the *B* positions within the oxygen octahedra of the spinel structure in a fully ordered way. The orthorhombic distortion results from a cooperative Jahn-Teller effect of the Cu$^{2+}$ ions at the octahedral sites. The CuO$_6$ octahedra form independent and infinite chains along the **b** direction leading to two nearly rectangular Cu-O-Cu super-exchange paths between nearest neighbor (NN) copper ions.[26] There is growing experimental and theoretical evidence that the title compound behaves like a one-dimensional $S$ = 1/2 Heisenberg AFM, although details of the next-nearest neighbor (NNN) exchange remain to be clarified.[25,27,28,29,30] Magnetic susceptibility, nuclear magnetic resonance (NMR), and ESR experiments revealed an average exchange coupling constant $J \sim 42$ K.[27,29] From neutron diffraction, long range magnetic ordering with a propagation vector **Q** = (0,0.53,0) has been observed below 2.1 K.[21] The incommensurate (IC) magnetic order is characterized by Cu$^{2+}$ moments which lie within the **ab**-plane with a pitch angle close to 90° and an ordered moment of 0.31 $\mu_B$. The dispersion of the magnetic excitations has been measured by inelastic neutron scattering. A detailed analysis allowed the determination of the relevant exchange paths resulting in NN ferromagnetic (FM) exchange ($J_1 \approx$ - 19 K), which is active via the two 90° Cu–O–Cu bonds and NNN AFM exchange ($J_2 \approx$ 45 K) acting via Cu–O–O–Cu super-exchange paths. Using quantum spin models[19] these parameters define a spiral spin ground state with a pitch angle close to 90°, as experimentally observed. A classical Hamiltonian would result in a much smaller turn angle between neighboring spins along the chain. This fact has been taken as proof for the importance of quantum fluctuations in LiCuVO$_4$.[22]

## II. EXPERIMENTAL DETAILS

LiCuVO$_4$ single crystals have been prepared as described in detail in Ref. 31. They crystallize in the space group *Imma* and reveal a transition to long-range magnetic order at $T_N$ = 2.5 K. The single crystals used in this work have been characterized by magnetic susceptibility, NMR, and antiferromagnetic resonance techniques.[23] Special care has been taken to choose crystals with almost ideal Li and Cu sublattices. NMR spectra were taken to exclude samples with a non-negligible amount of lithium and copper site-disorder.[23] The dielectric measurements were performed for electrical field directions along the three crystallographic axes. For this purpose, silver paint contacts were applied to the plate-like single crystals, either in sandwich geometry or by covering two opposite ends of the sample in a "cap"-like fashion thereby leaving a small rectangular gap. The complex dielectric constant was measured for frequencies between 320 Hz and 10 kHz using an Andeen-Hagerling AH2700A high-precision capacitance bridge. For measurements between 1.5 K and 300 K and in external magnetic fields up to 10 T a Quantum-Design Physical-Property-Measurement-System and an Oxford cryomagnet, equipped with a superconducting magnet were used. To probe ferroelectric order, we measured both the pyroelectric current at fixed magnetic field $H$, and the magnetoelectric current at fixed temperature using a high-precision electrometer. The current was integrated to determine the spontaneous polarization. To align the ferroelectric domains when cooling the sample through the transition temperature, we applied a polarizing field of the order of 1 kV/cm.

## III. RESULTS AND DISCUSSION

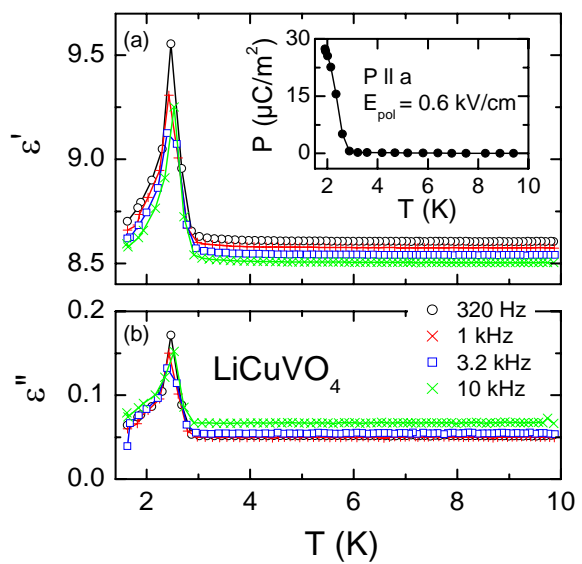

FIG. 1. (color online) Temperature dependence of the dielectric constant (a) and loss (b) of LiCuVO$_4$ at T ≤ 10 K for four frequencies with **E** || **a**. The inset shows the temperature dependent polarization along **a** direction, measured after polarizing the sample during cooling with an electric field of 0.6 kV/cm.

Fig. 1 shows the temperature dependence of the complex dielectric constant as measured for frequencies between 320 Hz and 10 kHz. The upper frame [Fig. 1(a)] documents the real part of the dielectric constant $\varepsilon'(T)$ around the antiferromagnetic phase transition, the lower frame (b) the dielectric loss $\varepsilon''(T)$. These experiments have been performed with the electric field **E** directed along the crystallographic **a** direction. Real and imaginary part have a very similar shape with a steep rise below 2.8 K, a maximum close to 2.5 K which corresponds to the magnetic phase transition and a somewhat smoother decrease towards low temperatures. The peak in $\varepsilon'$ signals a transition into a FE state. Neither $\varepsilon'$ nor $\varepsilon''$ reveal any significant frequency dependence. This seems natural, as LiCuVO$_4$ certainly has to be characterized as improper ferroelectric where long-range polar order is induced by the onset of spiral spin order. Hence no slowing down of polar relaxations is expected. These results are consistent with those published in Refs. 2 and 3, which however have been obtained at a single frequency only and without providing any information on the dielectric loss or the absolute values of $\varepsilon'$. In the inset of Fig. 1(a) we plot the spontaneous electrical polarization. It appears exactly at $T_N$ and increases to values of approximately 30 μC/m$^2$ at 1.9 K, similar to the observation reported in Ref. 3. A rough extrapolation of $P(T)$ towards 0 K gives a saturated polarization of approximately 50 μC/m$^2$, a value which is by a factor of 2 lower compared to the polarization in Ni$_3$V$_2$O$_8$ (Ref. 17) and by a factor of 10 lower when compared to the rare-earth manganites.[14] It is, however, by a factor of 10 higher when compared to the ferroelectric quantum spin chain LiCu$_2$O$_2$.[1]

The main aim of this work is to follow the dependence of the polarization as function of the external magnetic field and to check the symmetry relations for spiral magnets as predicted theoretically. In a detailed investigation, Park et al.[1] followed the evolution of the polarization $P(T,H)$ in LiCu$_2$O$_2$ along different crystallographic directions in various external magnetic fields. A clear interpretation of the results was hampered by the lack of knowledge about the complex spin configurations as function of temperature and magnetic field. In previous dielectric work on LiCuVO$_4$,[2,3] the temperature dependence of the capacitance **C** along the crystallographic **a** direction for various magnetic fields was investigated. In addition, the polarization $P(T)$ was measured for various external magnetic fields. However, a systematic study of the magnetodielectric phase diagram of LiCuVO$_4$ is still missing.

Recently the $(T,H)$ phase diagram of the magnetic phases in LiCuVO$_4$ has been determined by heat capacity, magnetic susceptibility, and high-field magnetization[24] as well as magnetic resonance studies.[23] The deduced spin configurations in the different magnetic states, obtained depending on external magnetic field at low temperatures, are schematically sketched in Fig. 2. At zero field these reports find a spin helix within the **ab**-plane (**e** ∥ **c**), with a pitch angle of 83.6° propagating in **b** direction (**Q** ∥ **b**). At a critical field $H_1 \approx 2.5$ T, the vector **e**, which is orthogonal to the spiral plane, is turned into the direction of the external field. Depending on the direction of this field, the normal vector **e** can point along the crystallographic **a**, **b**, or **c** direction, but the propagation vector **Q** remains unchanged. Finally, above a critical magnetic field $H_2$ the spiral structure is suppressed. From combined NMR and ESR experiments,[23] it has been concluded that the longitudinal spin component is modulated while the transverse component becomes disordered. We term this partly ordered state "modulated collinear" (see Fig. 2). However, the details of this magnetic structure have to be clarified by neutron scattering experiments. From this phase diagram it is immediately clear that the polarization should be switchable as function of the magnetic field and can be fully suppressed for fields $H > H_2$. This seems to be an ideal playground to test the proposed symmetry constraints for spiral magnets as outlined above.

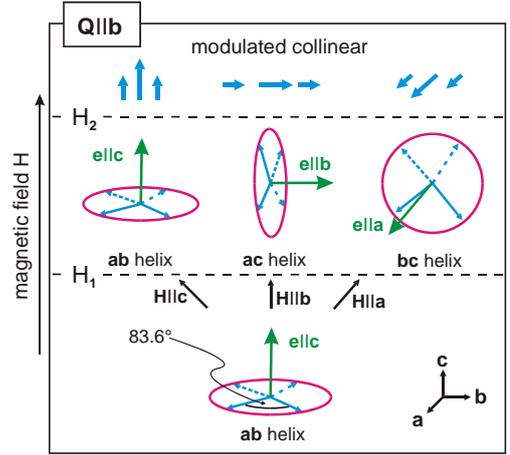

FIG. 2. (Color online) Schematic sketch of the spin configurations in dependence of external magnetic field in magnetically ordered LiCuVO$_4$ ($T < 2$ K).[23] $H_1$ and $H_2$ indicate phase boundaries between the different magnetic phases. The cone-like spin arrangement with a ferromagnetic component in the direction of the normal vector **e**, which likely appears between the fields $H_1$ and $H_2$, is not indicated.

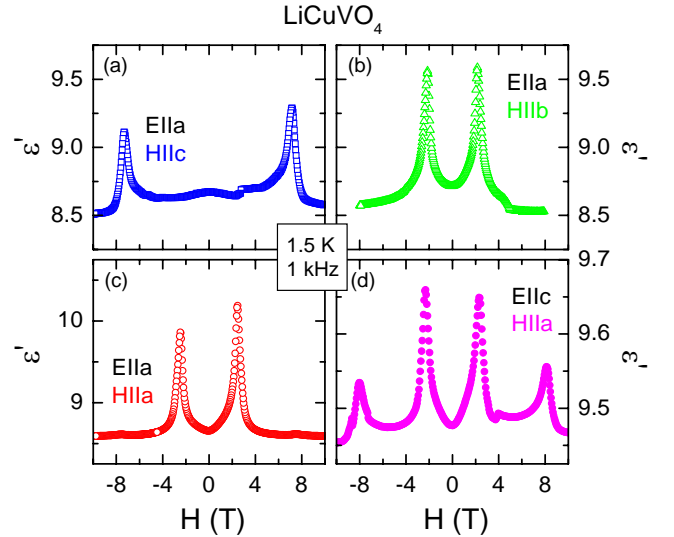

FIG. 3. (Color online) Magnetic-field dependent dielectric constant of LiCuVO$_4$ at 1 kHz and 1.5 K. The measurements were performed for different directions of electric and magnetic field as indicated in the figure. All data have been collected using field sweeps from +10 to -10 T.

One main result of this investigation is documented in Fig. 3. In zero external magnetic field, **Q** || **b** holds for the propagation vector and **e** || **c** for the normal vector of the spiral plane. In spiral magnets where the inverse Dzyaloshinskii-Moryia interaction or spin currents induce the electrical polarization, the polarization **P** is proportional to **e** × **Q**. Accordingly, we expect the ferroelectric polarization to be oriented along **a**, which indeed is observed (see Fig. 1). If we apply an external magnetic field along **c**, according to Fig. 2, when $H$ exceeds $H_1$ the vector **e** will remain parallel to **c** and ferroelectricity with **P** || **a** is maintained up to $H_2$ where a collinear structure is established. In this case, we do not expect a phase transition on passing $H_1$ but probably a spin canting with a concomitant ferromagnetic moment appears, which increases on increasing field. Fig. 3(a) showing $\varepsilon'(H)$ measured along **a** for **H** || **c** indeed reveals peaks at about ±7.3 T, i.e. at $H_2$ and not at $H_1$. Peaks in $\varepsilon'$ are commonly found for ferroelectric transitions and thus Fig. 3(a) indicates that FE is stable within these field limits.

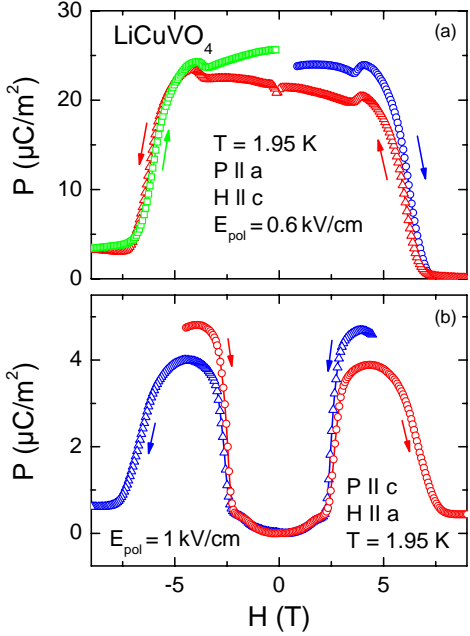

FIG. 4. (Color online) Magnetic-field dependent electrical polarization of LiCuVO$_4$ along **a** (a) and **c** direction (b). The measurements were performed at 1.95 K and for poling fields and magnetic-field directions as indicated in the figure. For the measurements of Fig. 4(a), the sample was cooled at 0 T with a subsequent measuring cycle as indicated by the arrows. The data of Fig. 4(b) were determined after cooling the sample with ±4 T.

If we instead apply the external magnetic field along the crystallographic **b** or **a** direction, at $H > H_1$ ferroelectricity is either completely suppressed (**H** || **b** → **e** || **b** → **P** ∝ **e** × **Q** = 0) or the polarization is turned into **c** direction (**H** || **a** → **e** || **a** → **P** ∝ **e** × **Q** ∝ **c**). Indeed, Figs. 3(b) and (c) reveal anomalies at about ±2.3 T, corresponding to the critical field $H_1$. Obviously, not only the ferroelectric-paraelectric transition [Fig. 3(b)], but also the mere change of the polarization direction leads to a peak in $\varepsilon'$ [Fig. 3(c)]. Now the critical test is a measurement that is sensitive to the polarization along **c** with the external magnetic field along **a**. At zero field, there is no polarization along the **c** direction, because for the spiral vector **e** || **c** holds and thus **P** ∝ **e** × **Q** is parallel to **a**. However, on increasing field, above $H_1$ the normal vector of the spiral reorients parallel to the external field **H** || **a**, yielding **e** || **a** and finite polarization along **c**. This FE state with **P** || **c** has to break down again at **H**$_2$ where a collinear spin structure is established. The experimental results documented in Fig. 3(d) indeed are fully consistent with these considerations: At about 2.3 T, FE with **P** || **c** appears and vanishes again close to 8.1 T, both transitions leading to peaks in $\varepsilon'$. Obviously, for **H** || **a** the phase boundary $H_2$ is shifted to a somewhat higher value (8.1 T) compared to 7.3 T for **H** || **c**. This seems to signal a significant anisotropy of **H**$_2$.

To further strengthen the validity of these symmetry constraints and to directly document the polarization switching by external magnetic fields, we also measured the magnetoelectric current with different combinations of electric and magnetic field directions. Two representative results are documented in Figs. 4(a) and (b). Fig. 4(a) shows the field dependence (**H** || **c**) of the polarization along **a**. In this direction the spiral phase reveals macroscopic polarization and this electrical polarization decays when entering into the collinear spin state. At approximately 7.5 T the spontaneous polarization has decayed completely. The spiral remains unaffected when passing into the intermediate spin state as the external magnetic field conserves the **ab** helix of the spiral state because **H** || **c** and thus **e** || **c**. In zero external magnetic field, the polarization along the crystallographic **c** direction is zero [Fig. 4(b)]. Increasing the external magnetic field along **a** induces ferroelectricity at ±2.5 T, which vanishes again for $H > 7$ T. In this case electrical polarization **P** || **c** is induced for $H > H_1$ because the normal vector of the spiral plane **e** is turned into the magnetic field direction (**H** || **a** and **e** || **a**). The question remains, why in this case the polarization is almost by a factor of eight lower when compared to the case **P** || **a** and **H** || **c**: This striking anisotropy of the polarization in LiCuVO$_4$ is in accord with recent density functional calculations including spin orbit coupling.[32] In this theory it has been shown that the ferroelectric polarization essentially originates from the spin-orbit coupling at the Cu sites, which, dependent on the spin orientation, can yield an asymmetric electron-density distribution around the oxygen ions. The resulting polarization **P** || **a** induced by the **ab** helix was calculated to be approximately six times larger than **P** || **c** for the **bc** helix, in very good

agreement with our observations. It only seems that the absolute values of the polarization are overestimated roughly by a factor of two. However, one has to take into account that the measurements presented here were performed at 0.75 $T_N$. Note that within this framework the experimentally observed polarization is fully explained in terms of the electronic charge distribution and, thus, ionic displacements seem to play only a minor role. Finally, we have to state that there is a severe discrepancy of our results documented in Fig. 4(b) compared to Ref. 3, where zero polarization has been reported for **H** || **a** and **P** || **c** at 0 T as well as for 4 T. This disagreement remains unexplained.

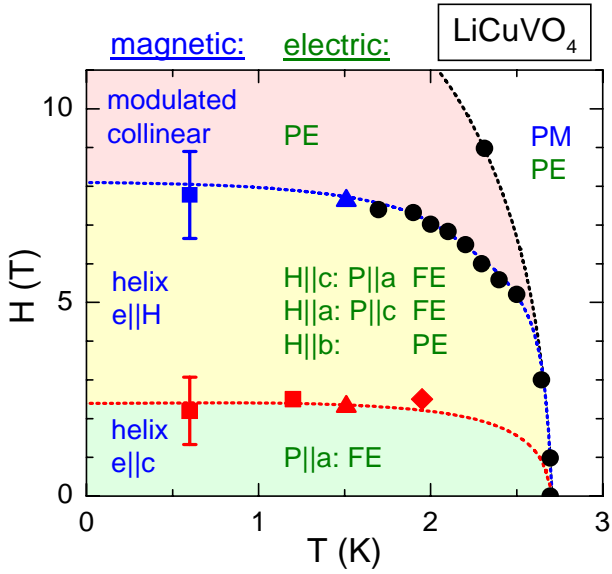

FIG. 5. (Color online) ($H$,$T$) phase diagram of LiCuVO$_4$. Results from the present work (triangles and lozenges) and from Refs. 24 (circles) and 23 (squares) were used. The magnetic (left column) and electric states (right column) are noted in the figure.

The results of Fig. 3 and 4 allow the construction of the magnetoelectric ($T$,$H$) phase-diagram shown in Fig. 5. In addition, results from Refs. 23 and 24 are included. In this schematic phase diagram we did not take into account the experimentally observed anisotropies of $H_2$. At low temperatures ($T < T_N \approx 2.5$ K) and at low external magnetic fields ($H < H_1 \approx 2.5$ T), LiCuVO$_4$ reveals helical spin order with a propagation vector **Q** along **b** and an (**a**,**b**) helix (normal vector **e** || **c**). According to the symmetry rule of spiral magnets, FE is established with the polarization **P** ∝ **e** × **Q** along the crystallographic **a** direction. On increasing magnetic field ($H_1 < H < H_2 \approx 7.5$ T) the normal vector **e** reorients along the external magnetic field and thus the electrical polarization depends on the direction of the magnetic field. In this regime the polarization can be switched from **a** to **c** direction, by turning the magnetic field from **c** to **a** direction. In these spin structures with **e** || **H** the propagation vector essentially remains the same as for $H < H_1$, i.e. **Q** || **b**. Hence, when the external magnetic field is along the crystallographic **b** direction, i.e. parallel to the propagation vector, LiCuVO$_4$ is paraelectric. Finally, for external magnetic fields above $H_2$ the helical spin structure becomes destroyed and the system is paraelectric for all field directions.

## IV. SUMMARY

In summary, we have performed a thorough characterization of the magnetocapacitive properties of multiferroic LiCuVO$_4$ by investigating the dielectric and polarization properties in dependence on temperature, magnetic field strength, and field direction. At 2.5 K this $S = 1/2$ quantum spin chain undergoes a transition into a helical spin structure with the spins rotating within the **ab** plane (normal vector **e** || **c**) and modulation **Q** || **b**. At 1.95 K the polarizability amounts 30 µC/m$^2$, considerably stronger than the polarization in the second known quantum spin chain LiCu$_2$O$_2$. We detected a considerable anisotropy of the polarization, with the absolute value of **P** strongly depending on the plane of the spin spiral, consistent with recent theoretical predictions.[32] Due to the ability to switch the direction of the **ab** helix in the helical spin-ordered state at $H_1 < H < H_2$, LiCuVO$_4$ is ideally suited to test the theoretically predicted symmetry relations for multiferroic spiral magnets.[9,10] We find an excellent agreement with these predictions and construct a detailed ($H$,$T$) phase diagram of this prototypical spiral-magnetic multiferroic.

## ACKNOWLEDGMENTS

This work was supported by the Deutsche Forschungsgemeinschaft via the Sonderforschungsbereich 484.

---

*Corresponding author. Email address: peter.lunkenheimer@Physik.Uni-Augsburg.de
†Present address: Universität zu Köln, II. Physikalisches Institut, 50937 Köln, Germany